\documentclass[11pt]{article}

\usepackage{achemso}

\usepackage{epsfig}
\usepackage{amsmath}
\usepackage{amssymb}
\usepackage{array}

\newcommand{\bfPhi}{\mathbf{\Phi}}
\newcommand{\bfx}{\mathbf{x}}

\begin{document}

\title{Resolution exchange  simulation with incremental coarsening}
\author{Edward Lyman\footnote{elyman@ccbb.pitt.edu} 
and Daniel M. Zuckerman\footnote{dmz@ccbb.pitt.edu}\\
Department of Computational Biology, School of Medicine\\
and Department of Environmental and Occupational Health,\\
Graduate School of Public Health,\\ BST W1041, 200 Lothrop St., University
of Pittsburgh, Pittsburgh, PA 15261}
\date{\today}
\maketitle
\begin{abstract}
We previously developed an algorithm, called ``resolution exchange'', 
which improves canonical sampling of atomic resolution models by   
swapping conformations between high- and low-resolution 
simulations\cite{resex05}. Here, we demonstrate a generally 
applicable incremental 
coarsening procedure and apply the algorithm to a larger 
peptide, met-enkephalin. In addition, we demonstrate a combination 
of resolution and temperature exchange, in which the coarser simulations 
are also at elevated temperatures. Both simulations are implemented 
in a ``top-down'' mode, to 
allow efficient allocation of CPU time among the different replicas.
\end{abstract}

Atomic resolution simulations of proteins are currently limited 
to short durations (less than  one $\mu$sec)\cite{schulten-f1-biophys04} or small 
systems (less than $100$ residues)\cite{duan-science98,pande-nature02}. Furthermore, accurate 
calculations involving large conformational changes 
are not possible for any system, as the cost of calculating 
entropic contributions is too great. Indeed, the cost of such 
calculations is only going to increase, as empirical force fields 
are improved by including polarization effects, either in a 
classical way\cite{halgren-polar-currop01,roux-polar-jctc2005} 
or in a semiclassical way\cite{tuckerman-carpar-pnas2005}. 

Thoroughly sampling the space of conformations is essential for 
a number of problems. From a purely biological perspective, there is a 
growing awareness of the importance of protein fluctuations---over and 
above the static picture---in the function of most 
proteins\cite{montelione-nat05}. Allostery 
and conformational changes dramatic enough to be captured experimentally 
are just two examples of the existence of such 
fluctuations\cite{gordon-glu-biochem2000,kern-allosteric-science01}. 
In a computational context, careful validation of empirical 
forcefields requires confidence in the quality of conformational 
sampling, so that error may be attributed to the forcefield rather 
than undersampling. The calculation of 
free energy differences---as required for evaluation of binding 
affinities of small molecules\cite{shoichet-nature}, or the strength of protein-protein 
interactions\cite{carlos-capri}---also requires reliable 
sampling\cite{kollman-free-chemrev1997,rodinger-currop-2005}. 

The undersampling (or ``quasi-ergodicity'') problem is widely recognized, 
and consequently there have  
been many attempts to improve upon standard simulation protocols. 
Methods which aim to generate a canonical distribution of conformations 
include multiple time-step methods\cite{berne-mts-md94,berne-mts-mc02}, 
nonlinear variable transformations\cite{tuckerman-tranny-prl02}, 
J-walking\cite{jwalk-jcp90}, inverse renormalization group 
approaches\cite{bai-nato01}, and adaptive resolution methods\cite{kremer-adaptive-2005}. 
The most widely used class of methods, 
however, comprises the generalized ensemble 
approaches\cite{okamoto-rev-biopol01,garcia-foldrev-currop03,deem-rev-pccp05}. Of 
the generalized ensemble approaches, perhaps the simplest and most 
popular is parallel tempering, in which a number 
of copies of the system are evolved in parallel at different 
temperatures\cite{swendsen-repex,nemoto-repex,hansmann-pt97,okamoto-repex-md}. 
Occasionally, configurations are swapped between neighboring replicas, 
presumably allowing the low temperature replica to access more 
configuration space via high temperature conformations. 

Numerous\cite{garcia-betahair-prot01,berne-repex-pnas02,cbrooks-repex-biophys03,okamoto-jcp04,cbrooks-repex-jmb04,garcia-helix-pnas04}, as well as 
extensive\cite{garcia-fold-pnas03,garcia-press-pnas05} parallel tempering 
simulations have been published, 
including some which claim to demonstrate the superior efficiency of the 
method over standard molecular dynamics (MD) 
simulation\cite{garcia-metenk-prot02,simmerling-bsheet-jmb05,duan-efficiency-jcp05}. 
Regardless of the validity of those claims, there appears to be  
a limit to the utility of parallel tempering for the study of large 
proteins, nucleic acids, and macromolecular complexes: the number of 
replicas required to bridge a specified temperature gap increases 
as the square root of the number of degrees of 
freedom of the solution\cite{kofke-accep-jcp02,kofke-accep}.  
In other words, if atomic resolution information is desired, then very many
atomic resolution simulations are required. Recent work by Berne and 
coworkers partly addresses this problem for explicitly solvated systems, so that the number of replicas scales 
with the number of degrees of freedom of the solute only\cite{berne-rest-pnas05}. Solutes 
like proteins, of course, can be quite large.     

In this paper, we address the problem of insufficient sampling of 
implicitly solvated biomolecules using a different approach. We 
recently developed an algorithm, called resolution exchange, 
which uses a distribution generated by a coarse-grained model  
to improve the sampling of a higher-resolution simulation\cite{resex05}. 
The resolution exchange (ResEx) algorithm guarantees canonical 
sampling for each level of resolution, so that the coarse-grained 
simulation introduces no bias into the high-resolution 
simulation. The algorithm is similiar in spirit to other exchange 
simulations, in that conformations are swapped between 
otherwise independent simulations. However, by employing replicas of reduced 
resolution, ResEx has the potential for significant efficiency gains. Other 
authors have recognized the value of improving sampling with reduced resolution 
representations. For example, Iftimie \emph{et.\ al.} used a classical potential 
as an importance function to improve sampling of an \emph{ab initio} 
potential\cite{iftimie-impsamp-qm-jcp2000}. Here, our goal is to sample a \emph{classical} 
atomic resolution potential, which leads us to a different algorithm. Also, 
Liu and Sabatti formally introduced a Markov chain Monte Carlo method which 
allows jumps between spaces of different dimensions\cite{liu-sintering-98}. 
Their algorithm has apparently 
not been applied to the simulation of macromolecules. 
Lwin and Luo recently introduced an algorithm similiar to ours, but it does not 
generate canonical sampling\cite{lwin-drem05}. 

We have also employed a modification of the usual parallel 
protocol used to carry out exchange simulations\cite{resex05}, 
generalizing the ``J-walking'' approach 
previously introduced by Frantz \emph{et.\ al.\ }\cite{jwalk-jcp90}. 
The J-walking (or as we call it, ``top-down'' exchange) method allows an 
unequal 
distribution of CPU time among the various replicas. We emphasize that any exchange 
simulation may be run in top-down mode. 
In contrast with other exchange methods, top-down exchange 
allows very little simulation time to be spent on the 
computationally expensive, high-resolution (or low temperature) replicas. 
Substantial improvement in sampling efficiency is therefore possible, in principle. 

We previously applied the resolution exchange algorithm to butane and  
dileucine peptide\cite{resex05}. Here, we confront issues which arise 
in the study of larger molecules. 
We show that a 
molecule may be coarsened incrementally, so that the overlap 
between models of neighboring resolution may be adjusted for 
improved sampling efficiency. We also demonstrate that resolution 
and temperature exchange are easily combined in a single simulation, so that sampling 
may be improved by both increasing temperature and decreasing resolution 
simultaneously. The incremental coarsening procedure is first demonstrated 
on dileucine, where we check that the correct conformational 
distribution is attained.   
We then demonstrate successful exchange between an all-atom 
model and an united-atom model of met-enkephalin, using two different exchange 
ladders: a ladder of varying resolution only, and a ladder which combines resolution 
and temperature changes. 
%
We will finish with a discussion of 
the next logical steps toward larger peptides and proteins. 
\section{Theory and methods}
The results presented in this paper concern two distinct, recently 
introduced simulation methods\cite{resex05}. 
The first is resolution exchange, 
which allows exchange between simulations 
at different resolutions, and preserves canonical sampling.
The second is 
top-down exchange, which allows unequal 
distribution of CPU time, maximizing the efficiency 
of an exchange simulation. In addition, we describe a general 
incremental coarsening strategy for building a ladder of models which 
improves exchange efficiency. 
\subsection{Resolution exchange\label{resexsec}}
Resolution exchange (ResEx) is motivated by the effectiveness of coarse-grained 
models for sampling of protein 
conformations\cite{warshel-nature,baker-fold-science05}, 
and by the need for atomic-level resolution for many calculations. Res-ex uses 
coarse-grained simulation to accelerate basin-hopping in more detailed 
models. In 
contrast with \emph{ad hoc} methods, ResEx guarantees canonical sampling 
of the atomic-resolution model.  

The basic idea, as in any exchange simulation, is to exchange conformations 
between two simulations. How are trial configurations constructed for  
an exchange between models with different numbers of degrees of freedom? 
Consider a pair of models: a coarse-grained model, a configuration of 
which is described by a set of coordinates $\mathbf{\Phi}$, and an atomic 
resolution model described by a larger set, $\{\mathbf{\Phi},\mathbf{x}\}$. 
Note that the coarse model is built from a \emph{subset} of the 
coordinates of the detailed model. Before the exchange, let 
the coarse-grained configuration be 
$\mathbf{\Phi}_{a}$, and let the atomic-resolution coordinates be 
$\{\mathbf{\Phi}_{b},\mathbf{x}_b\}$. By swapping only coarse variables, 
the trial configuration for the 
coarse-grained model is simply $\mathbf{\Phi}_{b}$, and for the 
atomic-resolution model is $\{\mathbf{\Phi}_{a},\mathbf{x}_b\}$. 

The exchange criterion is derived by considering the two simulations 
to consitute a single system characterized by the combined coordinates 
$\{\bfPhi_{a},(\bfPhi_{b},\bfx_{b})\}$. Because the simulations---aside from 
the exchanges---run independently, the probability distribution of the 
combined system is the simple product of the individual distributions. Let 
the potential functions of the high- and low-resolution simulations be 
$U_{H}(\bfPhi,\bfx)$ and $U_{L}(\bfPhi)$ respectively, and denote the Boltzmann factors as 
$\pi_{H}(\bfPhi,\bfx;\beta_{H})=e^{-\beta_{H}U_{H}(\bfPhi,\bfx)}/Z_{H}$ and 
$\pi_{L}(\bfPhi;\beta_{L})=e^{-\beta_{L}U_{L}(\bfPhi)}/Z_{L}$, where 
$Z_{H}$ and $Z_{L}$ are the partition functions. Then the exchange attempt is 
accepted with the Metropolis rate:
\begin{equation}
\text{min}\left [1,\frac{\pi_{H}(\bfPhi_{a},\bfx_{b};\beta_{H})}{\pi_{H}(\bfPhi_{b},\bfx_{b};\beta_{H})}\frac{\pi_{L}(\bfPhi_{b};\beta_{L})}{\pi_{L}(\bfPhi_{a};\beta_{L})}\right ].
\label{resxrate}
\end{equation}
The dependence on inverse temperature ($\beta$) is 
made explicit, as a reminder that the method is naturally combined 
with temperature exchange, though this of course extends to any type 
exchange, such as Hamiltonian exchange\cite{ham-repex}. Note that in the case 
of ordinary (temeprature based) replica exchange, in which all the coordinates are 
swapped, Eq.\ (\ref{resxrate}) reduces to the familiar expression 
$\text{min}[1,\text{exp}(-\Delta \beta \Delta U)]$.

In a parallel implementation, Eq.\ (\ref{resxrate}), together with the 
protocol for trial move construction, 
ensures that the algorithm satisifies the detailed balance 
condition. To see this, consider ``old'' ($o$) and trial/``new'' ($n$) 
configurations of the combined system. In 
the construction of any Boltzmann--preserving Monte Carlo move, two 
transition probabilites must be considered: the conditional probability $\alpha(o\rightarrow n)$ 
of \emph{generating} the move from configuration $o$ to $n$, and the conditional probability 
$w(o\rightarrow n)$ of \emph{accepting} the move\cite{frenkel-book}. Detailed balance insists that 
$p(o)\alpha(o\rightarrow n)w(o\rightarrow n)=p(n)\alpha(n\rightarrow o)w(n\rightarrow o)$, 
where $p(j)$ is the equilibrium probability of configuration $j$. 
Yet the acceptance criterion (\ref{resxrate}) has the form 
\begin{equation}
\frac{w\left(o\rightarrow n\right)}{w\left(n\rightarrow o\right)}=
\frac{p\left(n\right)}{p\left(o\right)},
\end{equation}
implying that the generating probabilities $\alpha$ are identical. This is indeed the case: 
given a pre-defined division into coarse and detailed coordinates, the conditional probability 
for the move  $o=\{\bfPhi_{a},(\bfPhi_{b},\bfx_{b})\}\rightarrow n=\{\bfPhi_{b},(\bfPhi_{a},\bfx_{b})\}$, 
and its inverse, are both one. That is, given the old configuration of the combined 
system, there is a unique trial configuration.
 
Lwin and Luo have constructed a similiar algorithm, 
except that \emph{before} checking acceptance via Eq.\ (\ref{resxrate}), the 
high-resolution trial configuration is \emph{minimized}\cite{lwin-drem05}. 
Such minimization (even subject to constraints on the coarse coordinates $\bfPhi$, as in 
ref.\ \cite{lwin-drem05}) violates the detailed balance condition 
by biasing the generating probability, $\alpha(o\rightarrow n)$, 
without any compensating correction in the acceptance criterion. 
Put more simply, reverse moves into un-minimized configurations are impossible. 
The consequences of the violation 
are readily seen, as shown in Sec.\ \ref{toymodel}. 

\subsection{Incremental Coarsening\label{incremental}}
An important practical issue is raised, however, by the construction of trial moves 
\emph{without} minimization. The problem is that the degrees of freedom in the high resolution 
simulation  
$\{\bfPhi,\bfx\}$ are strongly coupled---for a protein, think of $\bfPhi$ as backbone 
degrees of freedom (DoF) and $\bfx$ as side-chain DoF. Then it is clear that construction of 
trial moves by our method may lead to high rejection rates. We have solved 
this problem by noting that the rejection rate depends on the both the 
\emph{number and type} of DoF in the set $\{\bfx\}$. Employing a ladder of 
incremental models at intermediate resolutions allows the acceptance rates to be tuned to 
reasonable values, as shown in Fig.\ \ref{metenkladder}. 

A ladder of incrementally coarsened models is straightforward to construct. Consider 
coarsening from an all-atom representation of a protein to a united-atom 
representation. In the first model above the all-atom level, only one residue 
is described by the united-atom representation---the protein is described by 
a ``mixed model'', with one united-atom residue and the rest all-atom. Then, in the next 
level up, there are two united-atom residues, and so on, until the entire protein 
is described by the united-atom force field. A similiar procedure may then be used 
to go beyond the united-atom level to a united \emph{residue} level. Notice that it may 
be desirable to coarsen more than one residue at a time, since some residues 
have fewer degrees of freedom than others. 

Of course, implementation of the incremental ladder just described requires the 
construction of a potential function which has both united- and all-atom groups. In this 
work, we have built this mixed potential by combining the parameters for united and all-atom 
force fields into a single file. In other words, we created a larger parameter 
file, which contains both all-atom and united atom atom types. This file also 
includes all of the interactions for both united- and all-atom types, with the united-atom 
interactions modified as described in Sec.\ \ref{resexsec}. Adding some 
parameters (taken from the 
all-atom potential) for the interfaces, where united and 
all-atom residues link, the mixed potential describes the whole molecule. The 
parameters (formatted for use in TINKER) are included as supplementary material.   

The incremental coarsening approach just described is rather general and not restricted 
to implementing exchange ladders spanning united- to all-atom resolutions. Lower resolution 
models could also be considered, for which it may be desirable to coarsen several residues 
at once. A first quantitative analysis of the incremental coarsening procedure, 
suggesting how efficiency can be improved, is given in Sec.\ \ref{dileucine} and \ref{metenksection}. 
\subsection{Top-down exchange\label{topdownsec}}
In many exchange simulations, whether they are 
temperature-based\cite{duan-efficiency-jcp05}, 
Hamiltonian-based\cite{ham-repex}, or use some other extended 
ensemble\cite{depablo-hyper-jcp00,okamoto-multidim-repex}, the goal is 
to improve the sampling of a hard-to-sample model (such as an all-atom 
protein model at native conditions) by sampling a related model, 
which is \emph{presumed} easier-to-sample (such as the same all-atom model at 
higher temperature)\footnote{For a discussion of these issues from a 
statistical perspective, see Neal\cite{neal05}}. Information is 
swapped between the simulations by 
occasionally exchanging configurations, in a way which preserves 
canonical sampling of each distribution. Usually there is little 
overlap between the hard-to-sample (henceforth, ``bottom level'') 
and the easy-to-sample (henceforth, ``top-level'') models, 
and therefore a ladder of intermediate models is required.    

\begin{figure}[tp]
\epsfig{file=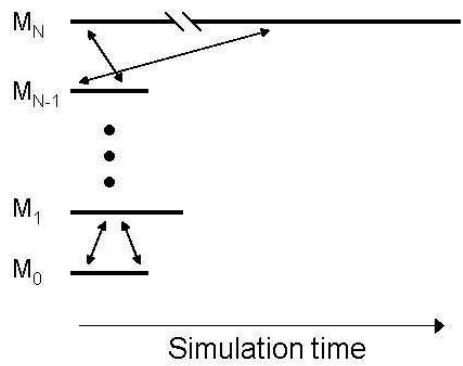,width = 0.5\columnwidth}
\caption{Schematic representation of top-down exchange. Thick horizontal 
lines are simulation trajectories (labelled ``M$_{i}$'' for model ``i'') 
and arrows represent exchanges. The M$_{i}$ may be differ in resolution, temperature,
or both. Notice that the top level simulation may 
be considerably longer than the others. }
\label{topdown}
\end{figure}

A critical observation is that the accuracy which is ultimately 
attained in the hard-to-sample, bottom-level model is effectively limited by that 
which is obtained in the easy-to-sample, top-level model\cite{jwalk-jcp90}. In many 
cases, the top level is still quite difficult to sample well, and will 
require considerable CPU time to reach an 
acceptable accuracy---much more than it would usually be allotted in a parallel implementation. 
It is this observation which motivates the top-down method.
The top-down algorithm shown schematically in Fig.\ \ref{topdown} was developed 
previously for temperature-based simulation\cite{jwalk-jcp90}, though was not widely recognized as such.  
The procedure is as follows:

\parbox{5in}{
(i) Run and store a simulation at the top level (model M$_{N}$) 
until it is judged to be 
sufficiently converged. This trajectory 
is a sample of the distribution $\pi_{N}(\mathbf{r})$ of the top level, 
where $N$ labels the level and $\mathbf{r}$ labels the configurations. 
In the case of ResEx, $\mathbf{r}= (\bfPhi,\bfx)$. Let $n$ be a running index, 
with $n=N$ at this top level.\\
(ii) Start a simulation at the $n-1$ level---for example, at the next lower 
temperature. Configurations $\mathbf{r}_{n-1}$ will be sampled according to 
$\pi_{n-1}$ for model M$_{n-1}$.\\
(iii) Whenever an exchange is to be attempted, pull a 
random trial configuration $\mathbf{r}_{n}$ from the M$_{n}$ trajectory. In the 
case of ResEx, one requires only the subset $\bfPhi$.\\ 
(iv) Accept the trial 
configuration according to\\
\begin{equation*}
\text{min}\left [1,\frac{\pi_{n-1}(\mathbf{r}_{n})}{\pi_{n-1}(\mathbf{r}_{n-1})}
\frac{\pi_{n}(\mathbf{r}_{n-1})}{\pi_{n}(\mathbf{r}_{n})}\right ],
\label{accept}
\end{equation*}
where $\pi_{i}(\mathbf{r})=\text{e}^{-\beta_{i}U_{i}(\mathbf{r})}/\text{Z}_{i}$, 
$\text{Z}_{i}$ is the partition function, $U_{i}$ is the potential function, 
and $\beta_{i}$ is the 
inverse temperature. Notice the partition functions need not be known, 
as they cancel between numerator and denominator.\\ 
(v) Continue with steps (iii) and (iv) until the sampling of the $n-1$ level 
is judged sufficient. Store the $n-1$ trajectory.\\
(vi) Continue with steps (ii) to (v) for $n=N-2,N-3,...$ until the bottom 
level simulation is complete.\\ 
}
 
First, note that canonical sampling is maintained 
by Eq.\ (\ref{accept})\cite{jwalk-jcp90}, just 
as in an ordinary parallel exchange simulation. On the other hand, detailed balance 
is not satisfied, as the trial configuration for 
the level $n$ simulation ($\mathbf{r}_{n-1}$ above) is discarded---making reverse 
moves effectively impossible. The error is one of 
practice, not of principle, arising from the fact that the samples of 
$\pi_{n}$ and $\pi_{n-1}$ are finite, just as in any simulation. 

To see intuitively that canonical sampling is maintained by top-down 
exchange, imagine a pair of simulations undergoing ordinary parallel 
exchange. Unbeknownst to the investigator, however, the top level simulation is 
running on a very fast processor, while the other is running on an old, 
slow processor. Between neighboring exchange points, the trial conformations from 
the fast processor will be far more decorrelated than those of the slow simulation---which mimics the effect of the top-down protocol. However, these 
exchanges still 
satisfy detailed balance. In the limit of an infinitely fast top-level simulation, 
trial configurations are completely decorrelated, and one could equally well choose 
randomly from $\pi_{n}$ as in step (iii). 

Second, notice that because trial configurations are 
pulled \emph{at random} from the sample of $\pi_{n}$ in step (iii), 
transitions which are slow in the actual M$_{n}$ trajectory 
occur rapidly among the trial configurations. Maximum benefit 
is thus obtained from successful exchanges---in contrast with a 
parallel exchange simulation, where trial configurations are typically 
separated by only a few picoseconds, and remain highly correlated.

Third, good results may be obtained expending very little CPU time on all 
levels except the top one. This may be understood from an energy landscape 
perspective. The top level has been used to thoroughly sample the 
space---high barriers are crossed, and major sub-basins equilibrated. 
At lower levels, only local equilibration need occur. For example, let 
$\tau_{nonloc}$ be the time to cross high barriers, $\tau_{local}$ be 
the time to equilibrate locally, and say that $m$ 
successful exchanges are needed to sample the space well. Then 
$\tau_{local}\times m$ CPU time is needed to sample the lower level. 
The required condition to save time over a parallel simulation is that 
$\tau_{local}<<\tau_{nonloc}$. The degree to which this condition 
is satisfied will depend on the system studied, but the top-down 
approach allows the flexibility to take advantage of a separation in 
time scales. This idea is reminiscient of the ``dragging'' of fast 
degrees of freedom, suggested by Neal\cite{neal-dragging04}, and the 
multiple time step approaches developed by Berne and coworkers\cite{berne-mts-md94,berne-mts-mc02}.

Finally, a major advantage of top-down simulation over parallel exchange protocols 
is that exchange attempts are  nearly ``free'', in the sense that no communication 
is required between processors\cite{jwalk-jcp90}. This means that exchanges 
may be attempted 
very frequently, and therefore much lower exchange rates may be 
accomodated. In the case of temperature exchange, this allows either for the steps in the 
temperature ladder to be more widely spaced, or for the treatment of larger systems 
with fewer replicas. 
\subsection{Simulation details\label{simdetails}}
In ideal circumstances, low-resolution models used in ResEx simulations would be 
specifically optimized for resolution exchange. They would have maximal conformational 
overlap for the common degrees of freedom. Here we work with an ``off the shelf'' low 
resolution model (united atom) which leads to some difficulties. 
Consider, for example, a C$^{\alpha}$--C$^{\prime}$ 
bond which is parameterized in the two models by two slightly different natural bond lengths. 
In an exchange attempt, the configurations are swapped, and in each trial configuration the 
C$^{\alpha}$--C$^{\prime}$ bond is moved a bit from its preferred length. These small contributions 
add up for every harmonic term in the entire molecule, and have a noticeable effect on the 
acceptance of exchange moves. We have solved this problem by simply changing the harmonic 
parameters of the coarse model to match those of the detailed model. This makes the coarse model 
more ``exchangeable'' with the detailed model. Since the coarse model is simply ``suggesting'' 
configurations for the atomic model, and since Eq.\ (\ref{resxrate}) guarantees that no bias is 
introduced by the coarse model, we need not worry that the coarse model parameters are changed from 
their original values. We now describe in detail the two molecular systems which were studied in the 
present work.
  
\emph{Dileucine. }We first studied dileucine peptide (ACE-(Leu)$_2$-NME) using the same forcefield 
parameters as in Ref.\ \cite{resex05}. 
Here, we also carried out an incrementally coarsened ResEx simulation of dileucine in 
$5$ levels. The coarsest level ($M_{4}$) was the same as in \cite{resex05}, namely a modified 
version of OPLSUA\cite{oplsua}. In lower levels, the molecule was rendered in finer detail 
beginning at the N-terminus: 
in $M_{3}$, the N-terminal methyl group, C$^{\alpha}$, and C$^{\beta}$ of the first residue 
were modelled in full atomistic detail; in $M_{2}$, C$^{\gamma}$ and both C$^{\delta}$'s 
of the first residue were additionally 
modelled explicitly; in $M_{1}$, the C$^{\alpha}$, C$^{\beta}$, C$^{\gamma}$, and 
one C$^{\delta}$ of the second residue were modelled explictly; and finally in $M_{0}$, 
the entire molecule was rendered in full atomic detail. The ladder of mixed models was chosen 
to keep approximately fixed the number of hydrogens by which neighboring levels differ, 
without splitting a methyl group. 

\begin{figure}[tp]
\epsfig{file=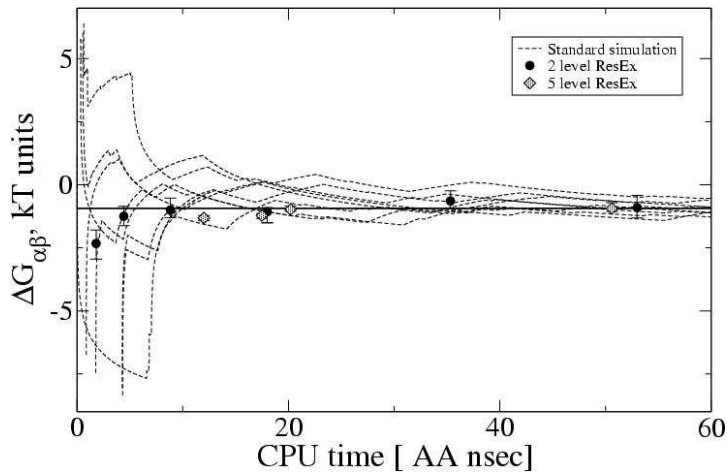,width = 0.75\columnwidth}
\caption{Comparison of different ResEx protocols for dileucine. Plotted 
are free energy difference estimates between the $\alpha$ and $\beta$ states 
as a function of total CPU cost. 
The dashed lines are individual runs generated by standard Langevin 
dynamics (no exchange), and the solid horizontal line is the avgerage of $4$ 
independent $150$ nsec Langevin dynamics simulations. The solid circles are 
ResEx results from the two level ladder from Ref.\ \cite{resex05}, 
and the diamonds are the ResEx results 
from the five level ladder, averaged in each case over $8$ independent runs. 
The error bars give the range of the $8$ independent runs. The ResEx data points 
are displaced from the origin to accurately reflect the time invested in generating 
the top level and all intermediate level distributions. The exchange schedule for ResEx 
has not been optimized. 
}
\label{dileufig}
\end{figure}

The top level ($M_{4}$) was simulated first, for $25$, $50$, $100$, or $200$ nsec. The 
different lengths of top-level simulation were used to generate the different data 
points in Fig.\ \ref{dileufig}. Then the higher resolution models were run, as per the top-down 
protocol (see Sec.\ \ref{topdownsec}), attempting exchanges once per psec. A total of $2.5\times10^3$ 
exchanges attempted between each level, for a total trajectory length per level 
of $2.5$ nsec. Frames were stored every $0.1$ psec, for a total of $2.5\times10^4$ frames 
in the sample at each level below the top.         

\emph{Met-enkephalin. }We next studied met-enkephalin (NH$_3^+$-Phe-Gly-Gly-Tyr-Met-COO$^-$). 
The united atom force field was a modified version of OPLSUA\cite{oplsua}.
The force 
field was modified so that the bond length and and angle bending 
parameters matched those of the all-atom force field, which improves 
exchangeability (or conformational overlap) of the two models. 
The sample of the top level 
model was constructed from two independent $100$ nsec trajectories, both 
started from pdb structure 1plw($1^{\text{st}}$ NMR model), generated by 
Langevin dynamics as implemented in TINKER v.\ $4.2$\cite{tinker}. The friction 
coefficient was $5$ psec$^{-1}$, and solvation was
modelled with the GB/SA method\cite{still-gbsa}. The first $1$ nsec of each trajectory 
was discarded and frames were stored every $10$ psec for a total of 
$19,800$ frames in the sample. 

We then ran the next higher resolution simulation, as per the top-down 
algorithm (see Sec.\ \ref{topdownsec}). This model was of mixed resolution, with the Tyr$_{1}$ 
residue represented by the OPLSAA all-atom forcefield\cite{oplsaa}, 
and the remaining residues described by the OPLSUA force field. Every $1$ psec, a 
random configuration was pulled from the top-level (M$_{5}$) trajectory, and a 
resolution exchange was attempted, with acceptance governed by Eq.\ (\ref{resxrate}). 
Since the acceptance ratio for the M$_{5}$ to M$_{4}$ 
exchanges was approximately $10$\%, the average length of M$_{4}$ trajectory between 
exchanges was $10$ psec. A total of $10^4$ ResEx 
moves were attempted, for a total M$_{4}$ trajectory length of $10$ nsec. 
Frames were stored every 
$0.1$ psec for a sample of $10^{5}$ frames. 

This procedure was then repeated 
for each level shown in Fig.\ \ref{metenkladder}, with the exception that the attempt frequency 
of ResEx moves was adjusted for the acceptance ratios, so that the segments of the simulations 
between exchanges were kept approximately constant at $10$ psec. Also, the total number of attempted 
exchanges was adjusted so that approximately $10^{3}$ successful exchanges were observed between 
each level, for a total trajectory length of $10$ nsec at each level. Given that the 
top level is presumed to be well-sampled, $10^{3}$ exchanges should sample a 
large number of basins.
\section{Results}
In a previous short paper, we tested the ResEx algorithm on two small molecules: butane 
and dileucine peptide\cite{resex05}. It was shown that the method produced results in 
agreement with those obtained by standard simulation methods. For the sake of clarity, here we 
first demonstrate our approach on a two-dimensional toy model, consisting of two basins which 
differ only \emph{entropically}. We also extend the method to two peptides, dileucine and 
met-enkephalin, in order to demonstrate the 
viability of incremental coarsening. 

\subsection{Results: Two-dimensional model\label{toymodel}}
\begin{figure}[tp]
\epsfig{file=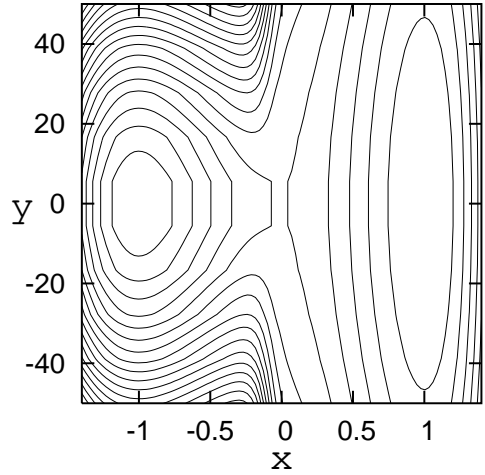,width = 0.75\columnwidth}
\caption{Contours of the potential surface $U(x,y)$, described by Eq.\ (\ref{Uxy}). Here we 
have reduced $w$ to $10$ so that both wells are visible in the figure.}
\label{twoDfig}
\end{figure}

An important consideration in designing any sampling method is whether it will 
correctly account for entropy differences. We therefore designed the potential surface shown in 
fig.\ \ref{twoDfig} to compare three 
different sampling methods: a ``standard'' Monte Carlo simulation,
the same Monte Carlo with resolution exchange, and the same Monte Carlo with the ``dual REM'' method
of Lwin and Luo\cite{lwin-drem05}. 

The surface $U(x,y)$ in fig.\ \ref{twoDfig} is described by the function 
\begin{equation}
U(x,y)=E_{b}\left( x^{2}-1\right)^{2} + \frac{E_{0}y^{2}}{1+w\left(\text{tanh}(x/0.1)+1\right)/2},
\label{Uxy}
\end{equation}
where $E_{0}\equiv k_{B}T$, $E_{b}=10$ $k_{B}T$ is the barrier height, and $w$ controls the 
width of the right well in the figure. Notice 
that the profile of the surface at $y=0$ is symmetric about 
$x=0$: $U_{x}(x)\equiv U(x,y=0)=E_{b}(x^{2}-1)^{2}$, i.e., the two minima are of equal 
depth.  The parameters were chosen so that the equilibrium populations of the two wells differ 
greatly---we used $w=500$, so that the right well holds $95$\% of the population, as measured 
by standard techniques. 

For both the ResEx and the dual REM simulations, the ``coarse-grained'' potential was simply the 
one-dimensional potential $U_{x}$, i.e., a symmetric double well. 

To describe the exchange moves explicitly, we denote $2D$ configurations by $(x,y)$ and 
$1D$ configurations by $\tilde{x}$. For both algorithms, an exchange 
move consists of two parts: the construction of a $1D$ trial configuration ($\tilde{x}_{new}$) 
from a $2D$ configuration 
($x_{old},y_{old}$), and vice versa: the construction of a $2D$ trial 
configuration ($x_{new},y_{new}$) from a 
$1D$ configuration ($\tilde{x}_{old}$). The construction of a $1D$ configuration in each 
case is simple--the ``extra'' 
($y$) coordinate is dropped, i.e., $\tilde{x}_{new}=x_{old}$. 

The only difference between the two simulations is in 
the construction of trial configurations for the $2D$ model from the $1D$ model. In ResEx, the trial 
configuration is the $\tilde{x}$ coordinate from the $1D$ model, with the (old) $y$ coordinate 
from the $2D$ model, 
i.e., $x_{new} = \tilde{x}_{old}$ and $y_{new} = y_{old}$. In dual REM, on the other hand, 
the trial $y$ coordinate 
is chosen randomly, and then minimized. For the potential $U(x,y)$, this 
means that $y_{new}=0$ \emph{always}, 
i.e., $x_{new} = \tilde{x}_{old}$ and $y_{new} = 0$. 

The ResEx simulation correctly samples the two wells, giving a population in the right well of 
$96.4\pm 1.6$\%. The dual REM simulation, on the other hand, yields a 
population of $51.0\pm 1.6$\% for the right 
well. What causes the error in the dual REM simulation? The answer is that the construction of dual REM 
trial moves violates detailed balance. More specifically, the minimization of the $y$ coordinate means 
that the difference in \emph{width} between the two wells is not accounted for correctly, since 
in dual REM $y_{new}=0$ \emph{always}. Notice that the it is not the random selection of the 
$y$ coordinate which intrinsically violates detailed balance, only the subsequent minimization. 

What is the analagous situation in molecular simulations? In this case, both ResEx and dual REM construct 
trial moves in internal coordinates--the coarse-grained model is built from a subset of the degrees of freedom 
of the atomic model. For example, the coarse-grained model (the $x$ coordinate above) may be the 
backbone coordinates of a protein, and the remainder (the $y$ coordinate above) may be the sidechain 
degrees of freedom. In dual REM construction of trial moves, the sidechains are minimized prior to 
exchange, and the therefore differences in entropy between different sidechain conformations are neglected. 
In ResEx, there is no minimization prior to exchange, and canonical sampling is maintained. 
\subsection{Results: Incremental coarsening of dileucine.\label{dileucine}}
We previously reported on ResEx results for dileucine, demonstrating successful exhange 
and significant speedup from a direct exchange between all-atom and united-atom models. Here, 
show that dileucine may be coarsened incrementally, and that (i) the correct distribution 
is observed for the all-atom model and (ii) adding additional intermediate levels of 
resolution improves efficiency. 

The additional levels boost the exchange acceptance by two orders of magnitude: exchanges were 
successful between $M_4$ and $M_3$ $15.5$\% of the time, between $M_3$ and $M_2$ $12.7$\% of the time, 
between $M_2$ and $M_1$ $29.0$\% of the time, and between and $M_1$ and $M_0$ $44.0$\% of the time. 
By comparison, exchanging AA and UA dileucine in a single step is successful 
only $0.16$\% of the time\cite{resex05}. However, we need to ask whether it is really more 
efficient to introduce additional levels of simulation in order to boost the acceptance 
of exchange moves. 

In fact, it appears to be substantially more efficient to use incremental coarsening rather than
abrupt coarsening. The cost for a given ladder of $N$ levels may be written
\begin{equation}
\text{total cost}=\text{top level cost}+\sum_{i=0}^{N-1}\frac{m\tau_{i}}{r_{i}},
\label{efficiency}
\end{equation}
where the cost of the top level is fixed, $m$ is the fixed number of
successful exchanges which are desired, $\tau_{i}$ is the simulation cost for
an interval between two exchange attempts at level $i$, and $r_{i}$ is the
acceptance rate between levels $i$ and $i+1$. We have assumed that
the sampling of level $i$ demands a fixed number of successful
resolution exchanges, consistent with the motivation of the top-down
protocol discussed in Sec.\ \ref{topdownsec}.

Eq.\ (\ref{efficiency}) implies that the effective exchange rate for an 
incremental ladder is a reciprocal sum of the individual rates. If we assume the 
$\tau_{i}$ are equal for all levels (which is exact for temperature exchange), then 
\begin{equation}
1/r_{\mathrm{eff}}=\sum_{i}1/r_{i},
\label{effrate}
\end{equation}
giving an effective rate for the $5$ level dileucine ladder of 
$5.1$\%. This result suggests an improvement in efficiency relative 
to the single step ladder, where the rate was $0.156$\%\cite{resex05}. 

In Fig.\ \ref{dileufig}, we compare the sampling of dileucine by three different 
simulation protocols: standard Langevin dynamics, resolution exchange with 
two levels, and resolution exchange with five levels. 
Sampling is assessed by examining the relative populations of the 
$\alpha$ and $\beta$ states ($\text{e}^{-\Delta\text{G}_{\alpha\beta}/k_{B}T}$) 
considered in Ref.\ \cite{resex05}. 
The convergence of this relative population measure requires transitions between the two states, 
which occur infrequently in a standard simulation. 
The five-level ladder clearly outperforms the two-level ladder, as we are able to 
generate results both more accurate and more precise with the five-level ladder 
in an equal amount of CPU time. Note that the total simulation time required for 
the entire ladder, including the top level, is included in the ResEx data points.

The efficacy of the ResEx approach is underscored by the fact that, at the top level 
(united atom), the sign of $\Delta\text{G}_{\alpha\beta}$ is wrong. That is, the exchange 
process corrects for a substantial bias in the coarse model.   
\subsection{Results: Incremental coarsening of met-enkephalin\label{metenksection}}
Met-enkephalin is a 
flexible neurotransmitter which participates in immune responses and 
pain inhibition, among other roles\cite{metenk-book,freed-metenk-solv}. 
By virtue of its small size and biological interest, it often is used to 
test new simulation methods\cite{okamoto-repex-md,remuca-bench-jcp03,repex-ris-jpcb04} 
and compare existing protocols\cite{freed-metenk-solv,freed-uavaa}.

Using met-enkephalin, we demonstrate the 
efficacy of the incremental 
coarsening procedure for a ladder of decreasing 
resolution at constant temperature, and for a ladder of simultaneously 
decreasing resolution and increasing temperature. 
Because quantifying the 
quality of sampling for met-enkephalin is considerably more difficult 
than is commonly appreciated, we will not present a detailed efficiency analysis. 
More will be said on this second topic in the 
discussion.

We employed the ResEx algorithm in a top-down framework, as sketched in 
Fig.\ \ref{topdown}. First, the top-level 
simulation (coarsest resolution---here, united-atom) was run. We then 
ran an exchange simulation at the next highest resolution---here, one residue was represented 
at all-atom resolution, and the rest of the peptide was united-atom. This 
procedure was continued, ``de-coarsening'' one residue at a time, until the entire peptide was 
represented at the all-atom level. Details are given in Secs.\ \ref{incremental} and \ref{simdetails}. 

\begin{figure}[tp]
\epsfig{file=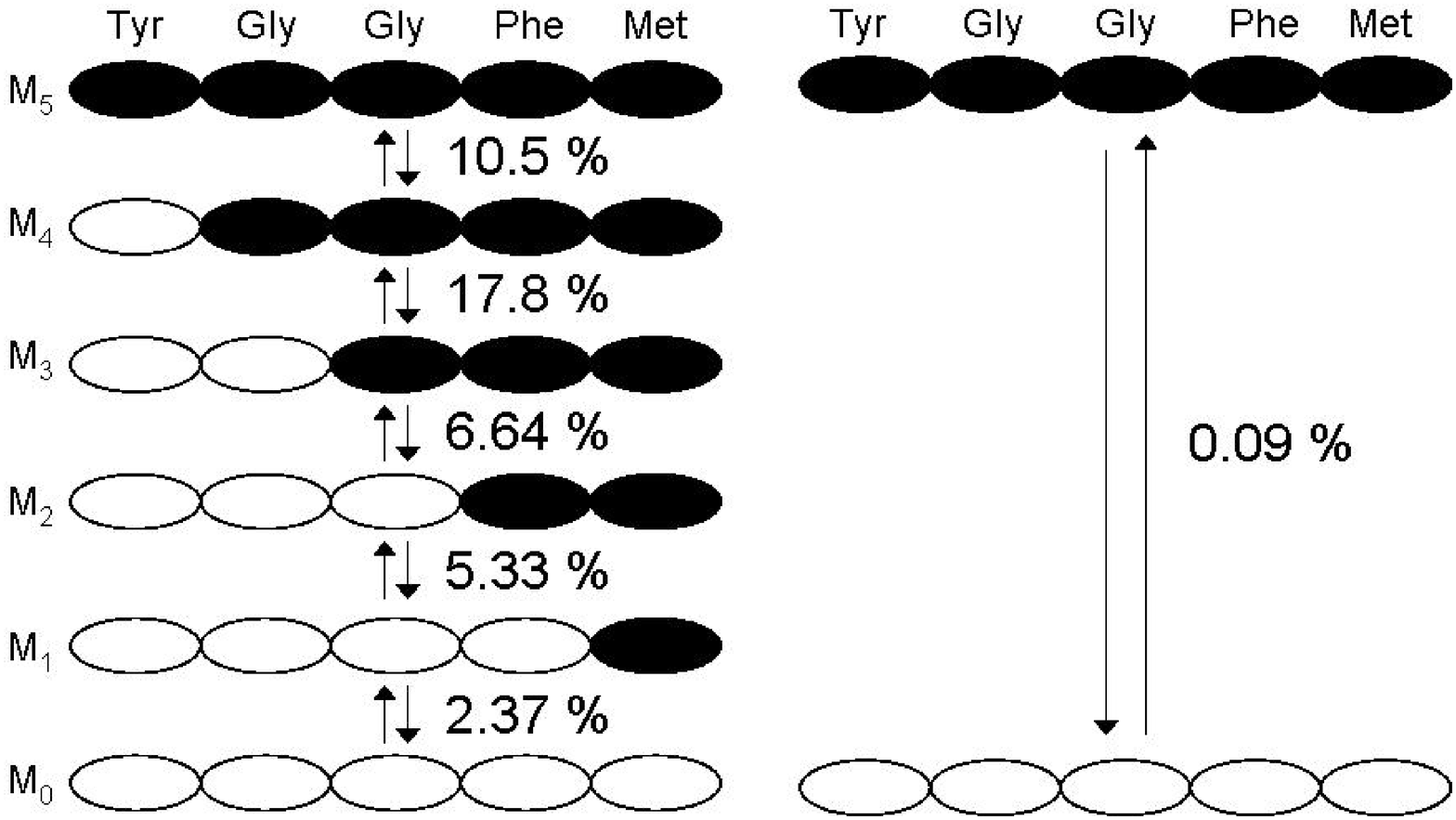,width = \columnwidth}
\caption{Two different ladders used to exchange all-atom with united-atom 
met-enkephalin. Residues are depicted with ovals---open corresponds 
to an all atom representation, filled to united atom. The ratios
of successful to attempted exchanges between each level are 
indicated by the percentages.}
\label{metenkladder}
\end{figure}

The incremental coarsening procedure substantially increases exchange acceptance. The 
five rates in the six-level ladder vary from $2.4$\% to $18$\%, as shown 
in Fig.\ \ref{metenkladder}. 
For comparison, exchanging between 
all-atom and united-atom models of met-enkephalin, with no intermediate 
levels of resolution, results in an acceptance ratio of $0.09$\%. 
The acceptance ratios vary, in part, according to the number of degrees of freedom 
by which the two levels differ. 


For met-enkephalin, the principal results are the acceptance rates shown 
in Fig.\ \ref{metenkladder}, which are significant for several reasons. 
First, they demonstrate the first implementation of the incremental 
coarsening approach in a complex peptide. Second, because they are well within the practical 
range of the top-down protocol---see Sec.\ \ref{topdownsec} and the 
Discussion---they indicate that the ResEx algorithm could prove important 
for larger peptides. Lastly, by comparing the effective exchange rate suggested 
by Eq.\ (\ref{effrate}), $r_{\mathrm{eff}}=5.2$\%, with the 
rate of $0.09$\% for direct exchanges between united- and all-atom models, 
one sees that a substantial speedup has been achieved. Of course, the 
magnitude of the improvement is merely suggestive---without a 
rigorous quantification of the sampling quality, there can be no rigorous 
comparison of efficiency. Such a quantification is beyond the scope of this 
work, as noted in the Discussion.  

It is useful to understand the intuitive reason behind the advantage of incremental coarsening.
If one writes the acceptance criteria (\ref{resxrate}) and (\ref{accept}) in 
the form $\text{min}[1,e^{-\epsilon}]$, then for exchanges between models 
of greatly differing resolution, one typically finds the dimensionless 
``energy'' is large, i.e, $\epsilon \gg 1$.  It seems to be roughly true that this energy is
proportional to the difference in the number of degrees of freedom in the models being
exchanged. However, if the change is made incrementally using many models M$_i$, then between
levels $i$ and $i+1$ there is a relatively small cost $\Delta \epsilon_i$, with
$\sum_i \Delta \epsilon_i \sim \epsilon$.  It is clear that with enough increments, one
can achieve $\Delta \epsilon_i \ll 1$, and thus create a high likelihood for exchange since
the corresponding Boltzmann factors are much larger: $r_i \sim e^{-\Delta \epsilon_i} 
\gg e^{-\epsilon}$.
This is exactly what is embodied in Eq.\ (\ref{efficiency}).  The trade-off is that one pays 
the cost for
simulating the additional intermediate levels.  However, as has been stressed in 
Sec.\ \ref{topdownsec}, the
intermediate-level simulations are very short compared to the top level.  In the present
context, for instance, the top level met-enkephalin trajectory is $198$ nsec, while all other levels were
simulated for only $10$ nsec.  The net savings can be quite substantial, especially considering 
that good sampling is achieved by increasing the number of exchanges.

\begin{figure}[bp]
\epsfig{file=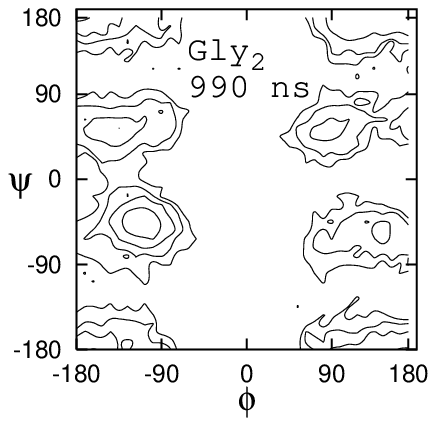,width = 1.5in}
\epsfig{file=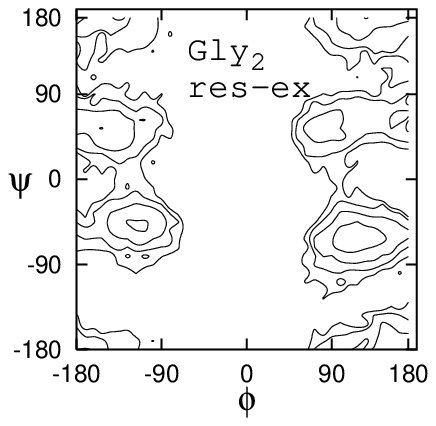,width = 1.5in}
\epsfig{file=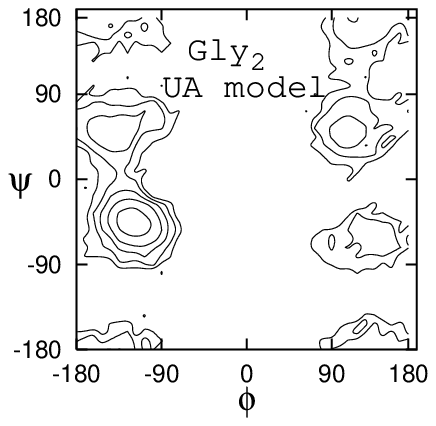,width = 1.5in}
\epsfig{file=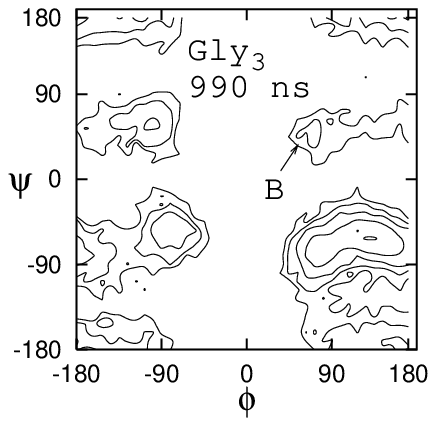,width = 1.5in}
\epsfig{file=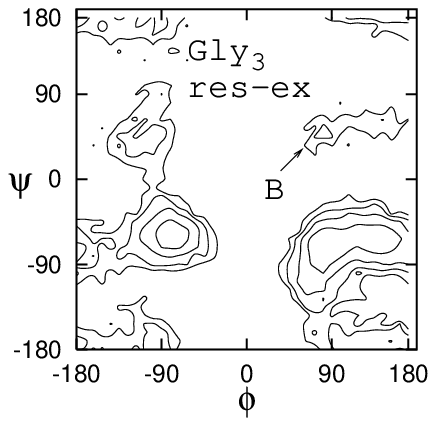,width = 1.5in}
\epsfig{file=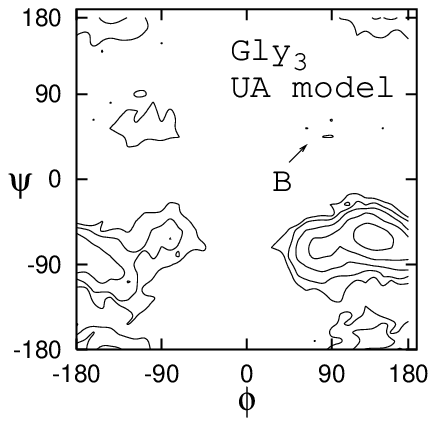,width = 1.5in}
\epsfig{file=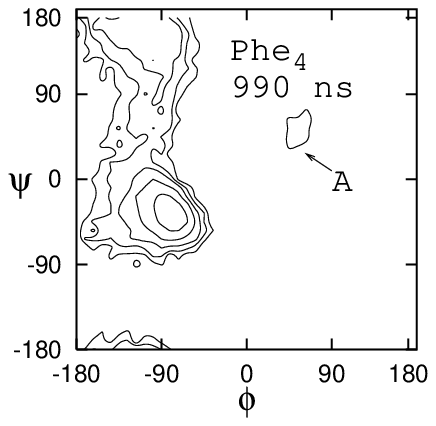,width = 1.5in}
\epsfig{file=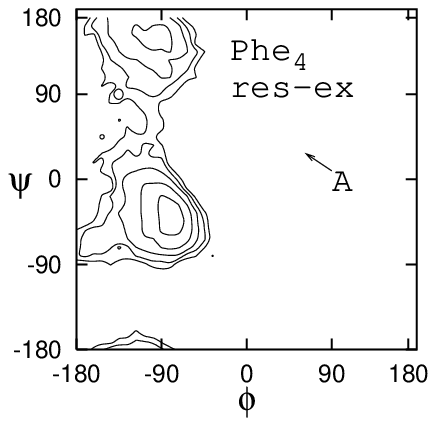,width = 1.5in}
\epsfig{file=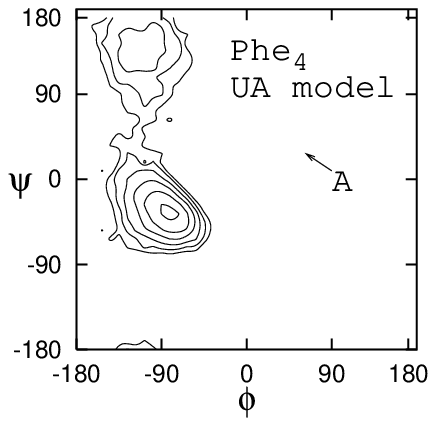,width = 1.5in}
\epsfig{file=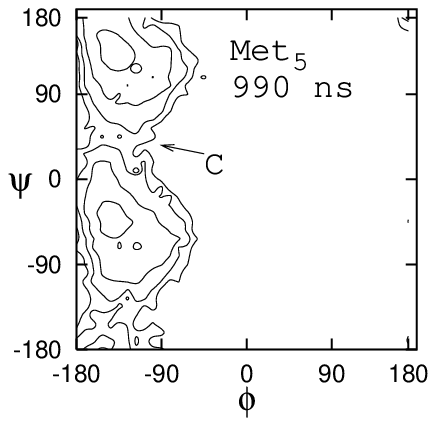,width = 1.5in}
\hspace{0.35cm}
\epsfig{file=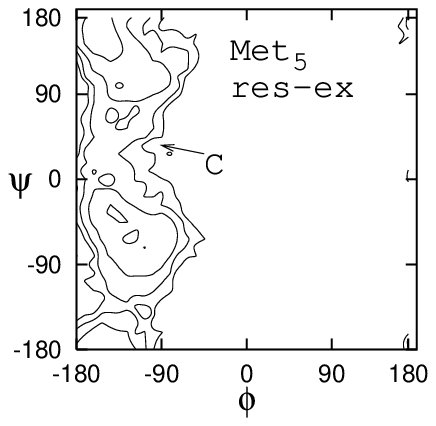,width = 1.5in}
\hspace{0.35cm}
\epsfig{file=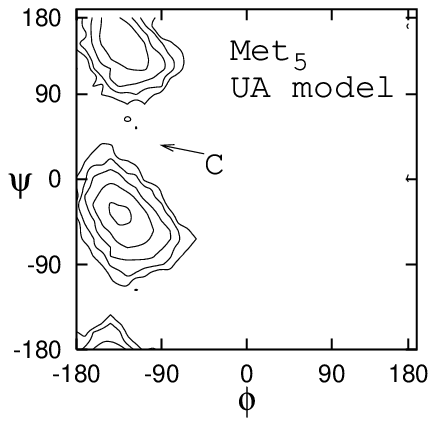,width = 1.5in}
\caption{Ramachandran histograms for met-enkephalin. The left column is 
a $990$ nsec Langevin dynamics simulation at all-atom resolution, 
without resolution exchange; the middle column is the all-atom level (M$_{0}$) 
from resolution exchange as described in the text; the right 
column is the top-level united-atom simulation (level M$_{5}$) used for the 
resolution exchange simulation shown in the middle column. Note that since the 
peptide is unblocked, there are only $4$ pairs of $\phi-\psi$ dihedrals. Res-ex fails 
to ``find'' one region (labelled ``A'') not present in the top-level simulation, but 
finds two others (labelled ``B'' and ``C''). 
}
\label{resex-rama}
\end{figure}

While we cannot yet rigorously measure sampling quality, we can show that 
the results obtained with ResEx are consistent with those obtained 
by standard methods, by comparing Ramachandran histograms (Fig.\ \ref{resex-rama}) 
from the ResEx simulation, to those obtained by standard simulation ($990$ nsec of 
simulation with the $M_{0}$ parameters). Overall, the agreement between the ResEx 
simulation and the $990$ nsec conventional simulation is quite good. However,  
a careful comparison reveals a region on the Phe$_{4}$ plot, labelled ``A'', which is noticeably 
under-sampled by the ResEx simulation, as compared to the $990$ nsec 
Langevin dynamics trajectory. The explanation is provided by an inspection of the Phe$_{4}$ 
histogram of the united-atom simulation: region ``A'' was not sampled by the top-level simulation. 
The failure points to a potential weakness of the 
ResEx (or any exchange) method---regions which are not sampled by the top level will be difficult to 
sample in any of the other levels. This is a specific instance of a general problem that occurs 
whenever auxiliary distributions are used to enhance sampling of some ``distribution of interest,'' 
namely the need to balance overlap with wider sampling via the auxiliary distribution\cite{neal05}.
In other words, it is a failure of the top-level simulation rather than the algorithm.

Interestingly, Fig.\ \ref{resex-rama} also presents two counterexmples to the forgoing discussion. Regions 
``B'' of the Gly$_{3}$ and ``C'' of the Met$_{5}$ plots were both well-sampled by the ResEx 
simulation, despite being infrequently visited by the top-level. That is, the ResEx 
acceptance criterion (\ref{resxrate}) correctly ``re-weights'' the conformation space of the 
all-atom model by allowing normal dynamics to continue when appropriate. Nevertheless, we are in 
the process of experimenting 
with other ``schedules'' 
(combinations of attempt frequency and number of exchange attempts) to balance the normal and the 
exchange dynamics. 

Ideally, the coarse model distribution would have better overlap with the 
high-resolution distribution, and the balance could be adjusted to favor 
exchanges over normal 
dynamics. This would allow the same quality of sampling with less simulation at each level 
below the top. In the long term, we hope to design coarse models constructed to \emph{not} 
eliminate \emph{any} regions of configuration space in more detailed models.   
%
%

\subsection{Resolution exchange with tempering}
We have also explored the possibilty of combining resolution exchange 
with parallel tempering, so that the sampling of the 
reduced models is improved both by the reduction in resolution and 
by increased temperatures. 
In a standard parallel 
tempering simulation, the temperatures are roughly exponentially 
distributed, in order that the conformational overlap between neighboring 
temperatures is constant over the ladder. However, there is no 
simple relationship between the change in \emph{resolution} and the 
acceptance of resolution exchanges. Some care must therefore 
be taken with the assignment of the temperature ladder. 

We began with the ladder of models in fig.\ \ref{metenkladder}. The 
acceptance ratios give an idea of the temperature gap which may be tolerated 
between two levels---a higher acceptance ratio will tolerate a larger jump 
in temperature. However, compared to standard parallel tempering 
simulations\cite{hansmann-pt97,okamoto-repex-md,garcia-betahair-prot01,berne-repex-pnas02,cbrooks-repex-biophys03,okamoto-jcp04,cbrooks-repex-jmb04,garcia-helix-pnas04,garcia-fold-pnas03,garcia-press-pnas05}, 
it may seem that the acceptance ratios are already 
too low to accomodate tempering in addition to resolution exchange. After all, 
we may expect that any difference in temperature will lower the acceptance 
of exchange moves. In this regard, the top-down approach has an important 
advantage over a parallel implementation. Since exchange attempts are 
``free'' (no commmunication between processors is required), they may be 
attempted much more frequently, and lower accptance ratios may be 
tolerated\cite{jwalk-jcp90}. 
Indeed, in our original study of dileucine peptide with top-down resolution 
exchange, the acceptance ratio was much less than $1$\%\cite{resex05}. 
See also Sec.\ \ref{topdownsec}.

\begin{figure}[bp]
\epsfig{file=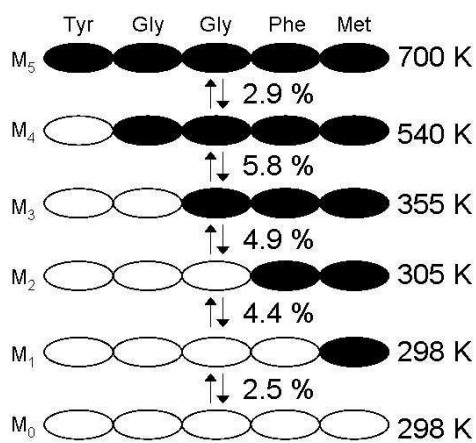,width = 0.5\columnwidth}
\caption{Ladder combining exchange between all-atom and united-atom 
met-enkephalin with tempering of reduced resolution simulations. 
Residues are depicted with ovals---open corresponds 
to an all atom representation, filled to united atom. The ratios
of successful to attempted exchanges between each level are 
indicated by the percentages. The temperature of each level is 
indicated on the right.}
\label{metenk-tres}
\end{figure}

The ladder combining temperature and resolution is shown in 
Fig.\ \ref{metenk-tres}. The temperature gaps were chosen by trial and 
error, aiming for an acceptance of attempted exchanges of a few 
percent between neighboring levels. Based upon this restriction, the top-level 
simulation was run at a temperature of $700$ K, which is comparable to 
previously published parallel tempering studies of 
met-enkephalin\cite{okamoto-repex-md,garcia-metenk-prot02}. We should expect, 
however, that fixed CPU cost sampling should be improved relative to ordinary replica exchange, 
by virtue of the reduction in resolution. 

The reduction in resolution confers an additional benefit when combined with 
tempering. Since the overlap between neighboring levels in a parallel tempering 
simulation scales like $(\text{number of DoF})^{1/2}$, reducing resolution allows 
the temperature gaps to \emph{increase} as the resolution is reduced. The 
overlap between neighboring levels in a combined resolution/tempering ladder is thus 
controlled both by the change in resolution, and the change in temperature, with 
the two effects compensating one another in an unknown way. Indeed, we observed one 
puzzling case in our search for an appropriate resolution/temperature ladder. 
In one ladder (data not shown), exchange 
between levels M$_{2}$ and M$_{3}$ was successful about $7$\% of the time when both were 
at $298$ K, while exchange occurred approximately $11$\% of the time when 
M$_{2}$ was thermostatted to $305$ K, and M$_{3}$ to $320$ K. We have not explained 
this result---though it should be remembered that different models have 
different landscapes, and therefore temperatures may not be directly compared.    
\section{Concluding Discussion}
We have extended our resolution exchange (ResEx) method\cite{resex05} using an 
incremental coarsening procedure for implicitly solvated peptides. After carefully 
testing the approach in the two-residue dileucine peptide, we applied it successfully 
to the five-residue met-enkephalin.
Incremental coarsening allows tuning of the 
conformational overlap between models of differing resolution, and therefore 
makes practical simulations which would otherwise be hampered by poor acceptance 
of exchange moves. We have also demonstrated that resolution exchange 
is naturally combined with parallel tempering, so that the reduced resolution models 
may be aided in their sampling of conformations by elevated temperatures. 

Ramachandran histograms demonstrate that, for the most part, ResEx simulation is consistent 
with standard simulation techniques. In one case, however, they reveal a weakness of our 
method---a top-level simulation which 
eliminates important regions of conformation space will result in poor sampling at the 
bottom level. This weakness is shared by any 
simulation which relies upon auxiliary ensembles to sample among major sub-basins. In 
the future we hope to eliminate this problem by more careful construction of 
reduced models.

Of course, we hope to treat still larger molecules with the ResEx method. 
Since it is essential that the top-level 
be well-sampled, the treatment of larger molecules will require yet coarser top-level 
simulation. This will likely require incremental coarsening from the united-atom level to 
a model with one or two beads per residue. Suitable models are under 
development. It appears that the ResEx approach cannot be applied easily to 
explicitly solvated systems; however, given the difficulty and 
importance of sampling implicitly solvated systems, ResEx may prove 
very valuable for biomolecular simulation.

We have also developed an alternative rigorous algorithm which permits the use of coarse top-level 
simulations to generate atomically detailed canonical samples. It is essentially a ``decorating'' 
procedure, and it eliminates the potential issue of correlations between coarse coordinates $\bfPhi$ 
and detailed coordinates $\bfx$, which could reduce acceptance rates in resolution exchange. 
Specifically, after generating a low-resolution sample distributed according to $\pi_{L}(\bfPhi)$, one can 
independently sample detailed coordinates $\bfx$ according to an arbitrary distribution $\pi_{x}(\bfx)$. 
(For example, $\pi_{x}$ could be based on harmonic terms in the full forcefield.) Full configurations are thus 
generated according to the simple product $\pi_{L}(\bfPhi)\pi_{x}(\bfx)$ and may be re-weighted to generate 
a fully detailed, high-resolution, distribution $\pi_{H}(\bfPhi,\bfx)$ using standard 
methods\cite{swendsen-singhist}. In the long term, the decorating approach 
may prove useful for adding explicit solvent. It may also be implemented in an  incremental 
fashion.  

An ``auxiliary'' question which remains to be carefully addressed is the quantification of sampling 
efficiency. There are numerous proposals for judging whether a simulation is converged---some are based 
on principal components\cite{hess-pre02}, others on energy-based ergodic 
measures\cite{thirumalai-jacs93}, and our own work in progress uses structural histograms\cite{el-dmz-class05}. 
Which one provides an appropriate measure depends on what properties are of interest. For 
applications which depend on the relative populations of various conformations, such as calculation of 
binding affinities for small molecules, a measure which depends directly on the 
conformational distribution is needed. Such a method is under development--for 
now we only mention that structural histograms provide a much more sensitive 
signal of non-convergence  than energy-based methods\cite{el-dmz-class05}.         

\emph{Acknowledgements.} The authors would like to thank Marty Ytreberg, Bruce Berne, Michael Deem, 
and Angel Garc\'{i}a for interesting discussions and insightful comments. This work was supported
by the Department of Environmental and Occupational Health and the Department of Computational
Biology at the University of Pittsburgh, and by the NIH (grants ES007318 and GM070987).  

\emph{Supporting Information.} Parameter files for the modified oplsua force field 
and for the mixed ua-aa force field are provided. The files are formatted for use 
with the Tinker molecular modelling simulation package. Also included is a sample 
`.key' file for running the mixed forcefield simulations in Tinker. 
This information is available 
free of charge via the internet at http:/pubs.acs.org.

\providecommand{\refin}[1]{\\ \textbf{Referenced in:} #1}

\clearpage

\clearpage
\clearpage
\clearpage
\clearpage
\clearpage


\end{document}